\documentclass[%
prd, twocolumn, nofootinbib, superscriptaddress]{revtex4-2}

\usepackage{aas_macros}
\usepackage{graphicx}
\usepackage{dcolumn}
\usepackage{bm}
\bmdefine{\bx}{x}
\bmdefine{\br}{r}
\newcommand{\phiobs}{\phi^\mathrm{obs}}
\usepackage{amsmath,amssymb}

\usepackage{color}

\begin{document}


\title{The Born approximation in wave optics gravitational lensing revisited}

\author{Hirotaka Yarimoto}
\affiliation{Department of Physics, Graduate School of Science, Chiba University, 1-33 Yayoicho, Inage, Chiba 263-8522, Japan}

\author{Masamune Oguri}
\affiliation{Center for Frontier Science, Chiba University, 1-33 Yayoicho, Inage, Chiba 263-8522, Japan}
\affiliation{Department of Physics, Graduate School of Science, Chiba University, 1-33 Yayoicho, Inage, Chiba 263-8522, Japan}

\date{\today}

\begin{abstract}
Gravitational lensing of gravitational waves provides us with much information about the Universe, including the dark matter distribution at small scales.
The information about lensed gravitational waves is encapsulated by an amplification factor, which is calculated by an integration of an oscillatory function. The Born approximation, which has been studied in terms of wave optics in gravitational lensing, may provide a means of overcoming the difficulty in evaluating the oscillating function and better understanding the connection between the amplification factor and the lens mass distribution. In this paper, we revisit the Born approximation for a single lens plane. We find that the distortion of gravitational waves induced by wave optics gravitational lensing is in general connected with the mass distribution of the lens object through a convolution integral, where the scale of the kernel is determined by the Fresnel scale. We then study the validity and accuracy of the Born approximation specifically for the case of a point mass lens for which the exact analytical expression of the amplification factor is available. Using the dimensionless parameter $y$, which represents the normalized impact parameter, and the dimensionless parameter $w$, which denotes the normalized frequency, we show that the $n$-th term of the Born approximation scales as $y^{-2}w^{n-1}$. This indicates that, for the case of a  point mass lens, the Born approximation is valid when  $w$ is less than 1, with its accuracy scaling as $wy^{-2}$ in this regime.

\end{abstract}

\maketitle


\section{Introduction}

Gravitational lensing is a powerful tool for investigating the Universe. Although the gravitational lensing effect has  mainly been applied to electromagnetic waves, it has become clear that gravitational waves are also subject to the gravitational lensing effect, just like electromagnetic waves (\cite{2019RPPh...82l6901O,1999PThPS.133..137N} for review articles). Gravitational wave lensing provides an important tool in unraveling the mysteries of the Universe. Examples include localizations of gravitational wave sources with sub-arcsecond precision (e.g., \cite{2020MNRAS.498.3395H}), cosmological parameter estimations (e.g.,  \cite{2003ApJ...595.1039T,2024PhRvD.109l4020C}), and searching for dark matter and compact objects (e.g., \cite{2023PhRvD.108j3529T,2022PhRvD.106b3018G}).

However, an amplification factor, which characterizes the extent to which gravitational waves are affected by the gravitational lensing effect, is expressed by an integral of an oscillation function and therefore is difficult to evaluate numerically unless the lensing object has a very simple mass distribution. Exact analytical forms of the amplification factor only exist for a limited number of mass density distributions such as a point mass lens\cite{1986ApJ...307...30D,2006JCAP...01..023M}. The series expansion also exists when the lens model is symmetric (e.g., \cite{2006JCAP...01..023M}), which is very costly to compute numerically in the high frequency or the massive lenses. We can obtain a lot of information about the Universe by analyzing the wave optic effect for a wide variety of lens mass distributions (e.g., \cite{2024arXiv240704052B,2022MNRAS.512....1G}). 

To calculate the amplification factor, it is useful to develop  methods that performs numerical computations efficiently  \cite{2024arXiv240904606V,2024PhRvD.109l4020C,2004A&A...423..787T,1995ApJ...442...67U,PhysRevD.102.124076}. Several approximations are sometimes used in such methods to reduce the computational cost. Understanding the validity and accuracy of the approximations in detail is crucial for improving the accuracy of the calculation of the amplification factor and reducing the computational cost further.

One of such approximations includes the so-called Born approximation \cite{2005A&A...438L...5T}. In the context of wave optics gravitational lensing, the Born approximation involves performing a successive expansion around freely propagating waves unperturbed by the gravitational potential and retaining terms up to the first order. 
In \citet{2006ApJ...644...80T}, it is shown that dispersions of the amplitude and phase of gravitational waves caused by gravitational lensing are connected to the power spectrum at small scale using the Born approximation. This is a clear benefit of using the Born approximation. In a more detailed study by \citet{2020ApJ...901...58O}, the detectability of dark low-mass halos and primordial black holes is explored to conclude that measuring gravitational lensing dispersions holds a significant potential for future gravitational wave observations (see also \cite{2021PhRvD.104f3001C,2023PhRvD.108j3529T,2023PhRvD.108j3532S}). 
Furthermore, \citet{2022PhRvD.106d3532O} shows that the amplitude and phase fluctuations are significantly enhanced by gravitational lensing magnifications due to geometrical optics lensing. Other applications include consistency relations between amplitude and phase fluctuations \cite{2021ApJ...918L..30I,2023PhRvD.108d4015T,2024PhRvD.109h3505M}.

In this paper, we focus on the Born approximation for a single lens plane. The accuracy of the Born approximation in the context of weak lensing by the large-scale structure has already been investigated in \citet{2023PhRvD.108d3511M} to complement \citet{2006ApJ...644...80T}. 
The investigation of the Born approximation for a single lens plane is important for e.g., lensing by a low-mass halo from which we want to extract information on the mass distribution of the halo. For this purpose it is of great importance to understand the connection between wave optics gravitational lensing observables and the mass distribution of the lens, which we explore in detail. Furthermore, using the Born approximation should help us evaluate the amplification factor because the phase in the integrand of the amplification factor becomes simpler in the Born approximation than without the approximation. However, the Born approximation including post-Born terms in gravitational lensing is not yet fully understood. It is one of our goals to deepen our understanding in this aspect.

\section{General properties of the Born approximation}

\subsection{Derivation}

First, we review the details of the Born approximation in wave optics gravitaional lensing. We consider scattering of gravitational waves caused by a weak gravitational field of a lens object. The background metric is the Friedmann-Robertson-Walker (FLRW) model with a weak gravitational potential 
$|\Phi(\bm{x})/c^2 |$ ($\ll 1$). The line elements are described by the metric $g_{\mu\nu}$ as $ds^2=g_{\mu\nu}dx^\mu dx^\nu=-(1+2\Phi/c^2)c^2dt^2+(1-2\Phi/c^2)dr^2$, which $\Phi$ is the gravitational potential and $dr^2 = a^2\chi^2(d\theta^2 + \sin^2\phi d\phi^2)$. We assume a flat universe and $\chi$ is radial comoving distance. Since the propagation equation for gravitational waves $\tilde{\phi}$ is described as $\partial_\mu\left(\sqrt{-g} g^{\mu \nu} \partial_\nu \tilde{\phi}\right)=0$, it is rewritten as \cite{1974PhRvD...9.2207P}
\begin{equation}
    \left[\bm{\nabla}^2+\left(\frac{\omega}{c}\right)^2\right]\phi(\bx)=\frac{4\omega^2}{c^4} \Phi(\bx)\phi(\bx),
    \label{eq:propagate}
\end{equation}
where $\tilde{\phi}(\bx)=\phi(\bx)e^{-i\omega t}$, $t$ is the conformal time , $\omega=2\pi f$ is the comoving frequency, $a\omega \rightarrow \omega$. In this set up, the scale factor $a$ is absorbed into the comoving frequency and does not appear explicitly. $\nabla$ is the three-dimensional Laplace operator. Let $\phi_0$ be the solution of equation (\ref{eq:propagate}) when there is no lensing effect ($\Phi=0$), and $\delta\phi$ be the term representing the effect of scattering caused by the gravitational lensing, we can describe $\phi$ as
\begin{align}
    \phi(\bx) &= \phi_0(\bx)+\delta \phi (\bm{x})  \nonumber\\
    &= \phi_0(\bx)-\frac{\omega^2}{\pi c^4}\int d^3\bm{x}' \frac{e^{i\omega|\bx-\bx '|/c}}{|\bx-\bx '|}\Phi(\bx ')\phi(\bx ').
    \label{eq:01}
\end{align}
The second line calculation uses the Green function of the Helmholtz equation $G(\bx ,\bx ')=-e^{i\omega |\bx-\bx '|/c}/4\pi|\bx - \bx '|$. Note that equation (\ref{eq:01}) is not the solution but an integral equation. We substitute equation (\ref{eq:01}) successively into equation (\ref{eq:01}) to obtain
\begin{align}
   \phi(\bx)=& \phi_0(\bx)-\frac{\omega^2}{\pi c^4}\int d^3\bm{x}'\frac{e^{i\omega|\bx-\bx '|/c}}{|\bx-\bx '|}\Phi(\bx ')\phi_0(\bx ') \nonumber\\ 
   &+\frac{\omega^4}{\pi^2 c^8}\int d^3\bm{x}'d^3\bm{x}''\frac{e^{i\omega|\bx-\bx '|/c}}{|\bx-\bx '|}\frac{e^{i\omega|\bx '-\bx ''|/c}}{|\bx '-\bx ''|}\nonumber\\
   &\times \Phi(\bx ') \Phi(\bx '')\phi_0(\bx '')-\cdots  \label{eq:afsuc}.
\end{align}   
 By taking the first and second terms of equation (\ref{eq:afsuc}) ($ |\phi_1| \ll |\phi_0| $)
 \begin{align}
  \phi(\bm{x}) &= \phi_0(\bm{x}) + \phi_1(\bm{x}) + \phi_2(\bm{x}) + \cdots \nonumber\\
  &\simeq \phi_0(\bm{x}) + \phi_1(\bm{x}), 
 \end{align}
  we obtain
 \begin{equation}
 \phi(\bx)= \phi_0(\bx)-\frac{\omega^2}{\pi c^4}\int {d^3\bm{x}'}\frac{e^{i\omega|\bx-\bx '|/c}}{|\bx-\bx '|}\Phi(\bx ')\phi_0(\bx '). \label{eq:born}
 \end{equation}
 This approximation is called the Born approximation \cite{2005A&A...438L...5T}. 
 
 The ratio of  $\phiobs_1$ to $\phiobs$ can be approximated as
 \begin{equation}
 \frac{\phiobs_1}{\phiobs} = \frac{\phiobs_1}{\phiobs_0+\phiobs_1} = \frac{\phiobs_1}{\phiobs_0(1+\frac{\phiobs_1}{\phiobs_0})} \simeq \frac{\phiobs_1}{\phiobs_0}, 
 \end{equation}
by ignoring the second order of $\phiobs_1/\phiobs_0$. We take the spherical wave $Ae^{i\omega|\bx-\bx_s|/c}/|\bx-\bx_s|$, where $\bm{x}_\text{s}$ is the position of the source object as the solution of $\phi_0$, and define $\phiobs$ as the observed wave at the origin ($\bx=\bm{0}$). Plugging $\phi_0$ into equation (\ref{eq:born}), we obtain
  \begin{multline}
  \frac{\phiobs_1}{\phiobs} \simeq \frac{\phiobs_1}{\phiobs_0} = -\frac{|\bx_s|}{Ae^{i\omega |\bx_s|/c}}\cdot \frac{\omega ^2}{\pi c^4} \\
  \times \int d^3\bm{x}' \frac{e^{i\omega |\bx '|}}{\bx '} \frac{Ae^{i\omega |\bx ' - \bx_s|/c}}{|\bx '-\bx_s|}\Phi (\bx '). \label{eq:02}
  \end{multline}
  
  We set the two-dimensional vector perpendicular to the line of sight as $\bm{r}$, i.e., $\bx =(\chi,\bm{r})$. Furthermore, we assume $|\bm{r}| \ll \chi$. Then we can approximate
  \begin{equation}
  |\bx-\bx_s| \simeq \chi_s - \chi + \frac{\chi_s \chi}{2(\chi_s-\chi)}\left|\frac{\bm{r}}{\chi}-\frac{\bm{r}_s}{\chi_s}\right|^2. 
  \end{equation}
  Using this approximation, we rewrite equation (\ref{eq:02}) as
  \begin{equation}
  \frac{\phiobs_1}{\phiobs_0} =  -\frac{\omega^2}{\pi c^4} \int d^3\bm{x}' \frac{\chi_s}{\chi'(\chi_s-\chi')}\Phi(\bx ')e^{i\omega \Delta t(\chi',\bm{r})} \label{eq:before} , 
  \end{equation}
      where
  \begin{align}
  \Delta t(\chi , \bm{r}) &= \frac{\chi_\text{s}}{2c\chi(\chi_\text{s}-\chi)}\left| \bm{r} - \frac{\chi}{\chi_\text{s}} \bm{r}_\text{s} \right|^2 \nonumber\\
  &= \frac{\chi_\text{s}}{2c\chi(\chi_\text{s}-\chi)} |\bm{r} - \bm{r}_\perp|^2 ,
  \end{align}
 with $\bm{r}_\perp = (\chi/\chi_\text{s}) \bm{r}_\text{s}$. We apply the thin lens approximation, for which the gravitational potential is written as
  \begin{equation}
  \Phi(\bx ') \simeq \frac{c^2}{2}\frac{\chi_s}{\chi(\chi-\chi_\text{s})}\delta^\text{D} (\chi '-\chi)\psi(\bm{r}) \label{eq:thin},
  \end{equation}
  where the gravitational lens potential $\psi$ is defined as
  \begin{equation}
  \psi(\bm{r}) \equiv  \frac{2}{c^2}\frac{\chi(\chi_s-\chi)}{\chi_s} \int d\chi ' \Phi(\chi ',\bm{r}).
  \end{equation}
  We substitute equation (\ref{eq:thin}) into equation (\ref{eq:before}) to obtain
  \begin{equation}
  \frac{\phiobs_1}{\phiobs_0} = -\frac{\omega^2}{2\pi c^2}\left(\frac{\chi_\text{s}}{\chi(\chi_\text{s}-\chi)}\right)^2 \int d^2\bm{r} \psi(\bm{r}) e^{i\omega \Delta t(\bm{r})} \label{eq:lensborn}. \\
  \end{equation}
   Note that $\chi$ represents the comiving radial distant to the lens object. 
   
   For the sake of convenience, we introduce the Fresnel scale \cite{2004A&A...422..761M} 
  \begin{equation}
  r_\mathrm{F} = \sqrt{\frac{c\chi(\chi_\text{s}-\chi)}{\omega \chi_\text{s}}},
  \end{equation}
  which is defined in comiving units. Using the Fresnel scale, equation (\ref{eq:lensborn}) is rewritten as
  \begin{equation}
    \frac{\phiobs_1}{\phiobs_0} = -\frac{1}{2\pi r_\text{F}^4} \int d^2\bm{r}  \psi(\bm{r}) \exp \left[ {i \frac{1}{2 r_\text{F}^2}|\bm{r} -\bm{r}_\perp|^2} \right]   \label{eq:fresnelborn}  .
   \end{equation}
  
  \subsection{Re-definition of the phase}\label{sec:redef_phase}
  
  Now we consider the geometrical optics limit, $f \rightarrow \infty $, or equivalently $r_\text{F} \rightarrow 0$. In this situation, the phase of the integrand of equation (\ref{eq:fresnelborn}) is so large that it oscillates violently. Thus, only the extreme values of the phase contribute to the integral. This is called the stationary phase approximation. In equation (\ref{eq:fresnelborn}) the extreme value is at $\bm{r} = \bm{r}_\perp$
  at which we assume $\psi(\bm{r})$ does not change dramatically. Equation (\ref{eq:fresnelborn}) is then rewritten as 
  \begin{align}
  \frac{\phiobs_1}{\phiobs_0} &\simeq   -\frac{1}{2\pi r_\text{F}^4} \psi(\bm{r}_\perp)  \int d^2\bm{r}  \exp \left[ {i \frac{1}{2 r_\text{F}^2}|\bm{r} -\bm{r}_\perp|^2} \right]  \nonumber \\
  &= -\frac{i}{r^2_\text{F}} \psi(\bm{r}_\perp),
   \label{eq:geo1}
  \end{align}
  where we used the following integral
  \begin{equation}
  \int ^\infty _0 e^{i \lambda u^p} du = e^{i\frac{\pi}{2p}} \lambda^{-\frac{1}{p}} \Gamma\left(\frac{1}{p}+1\right).
  \end{equation}
  For $r_\text{F}\rightarrow 0$, $\phiobs_1/\phiobs_0$ diverges in geometrical optic limit. Thus we impose the condition $\psi(\bm{r}_\perp) = 0$ by using the uncertainty of the potential with respect to a constant shift. Using this condition, equation (\ref{eq:fresnelborn}) can also be rewritten as
 \begin{align}
 \frac{\phiobs_1}{\phiobs_0} 
 &= -\frac{1}{2\pi r_\text{F}^4} \int d^2\bm{r} \psi(\bm{r})  \exp \left[ {i \frac{1}{2 r_\text{F}^2}|\bm{r} -\bm{r}_\perp|^2} \right]  \nonumber\\
 &\qquad -  \left(-\frac{i}{r^2_\text{F}} \psi(\bm{r}_\perp) \right)  \nonumber \\
 &=  -\frac{1}{2\pi r_\text{F}^4} \int d^2\bm{r}  \psi(\bm{r}) \nonumber\\
 &\qquad \times \left\{ \exp \left[ {i \frac{1}{2 r_\text{F}^2}|\bm{r} -\bm{r}_\perp|^2} \right] -2\pi i r^2_\text{F} \delta(\bm{r}-\bm{r}_\perp) \right\} .
 \label{eq:geo2}
 \end{align}
  
  \subsection{Born approximation in Fourier space}
  
\begin{figure}[tbp]
\includegraphics[width=9cm]{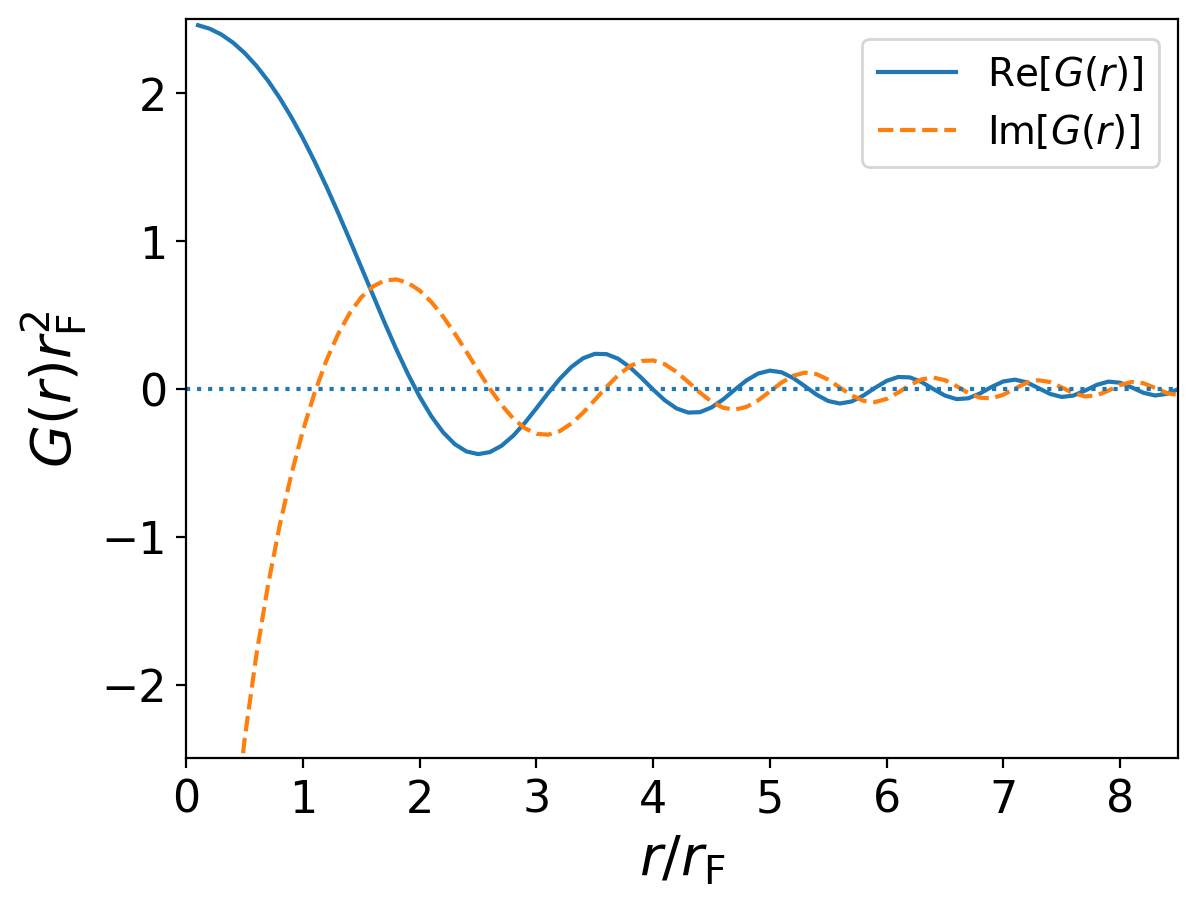}
\caption{\label{fig:epsart} The kernel function of the convolution defined by equation (\ref{eq:eq04}) as a function of the lens plane distance normalized by the Fresnel radius on a lens plane. The blue solid line represents the real part of the kernel and the orange dashed line represents the imaginary part of the kernel.  }
\end{figure}

  We use Fourier transform of $\psi(\bm{r})$ and $\kappa(\bm{r})$, defined as
  \begin{gather}
  \psi (\bm{r})= \int \frac{d^2\bm{k}}{(2\pi)^2} \tilde{\psi}(\bm{k}) e^{i\bm{k \cdot r}} \label{eq:fourierp}, \\
  \kappa(\bm{r}) = \int \frac{d^2\bm{k}}{(2\pi)^2} \tilde{\kappa}(\bm{k}) e^{i\bm{k \cdot r}}, \label{eq:fourierk}  
  \end{gather}
  which are related with each other in real space as 
  \begin{equation}
  \bm{\nabla}_{\bm{r}}^2 \psi(\bm{r}) = 2 \kappa(\bm{r}),
  \end{equation}
and in Fourier space as
   \begin{equation}
  -k^2 \tilde{\psi}(\bm{k}) = 2 \tilde{\kappa}(\bm{k})\label{eq:fourierpk} .
  \end{equation}
  By using equations (\ref{eq:fourierp}), (\ref{eq:fourierk}), and  (\ref{eq:fourierpk}), equation (\ref{eq:geo2}) is computed as
  \begin{align}
  \frac{\phiobs_1}{\phiobs_0} 
  =& -\frac{1}{2\pi r_\text{F}^4} \int d^2\bm{r}  \int \frac{d^2\bm{k}}{(2\pi)^2} \tilde{\psi}(\bm{k}) \nonumber\\
  & \times \Bigg\{ \exp \left[ {i \frac{1}{2 r_\text{F}^2}|\bm{r} -\bm{r}_\perp|^2 + i\bm{k} \cdot \bm{r}} \right] \nonumber\\
  &-2\pi i r^2_\text{F}  e^{i\bm{k} \cdot \bm{r} }\delta(\bm{r}-\bm{r}_\perp) \Bigg\},
  \label{eq:eq03}
  \end{align}
  and the exponent in equation (\ref{eq:eq03}) reduces to
\begin{align}
i \frac{1}{2 r_{\mathrm{F}}^2}\left|\boldsymbol{r}-\boldsymbol{r}_{\perp}\right|^2+i \boldsymbol{k} \cdot \boldsymbol{r}=&
i \frac{1}{2 r_{\mathrm{F}}^2}\left|\boldsymbol{r}-\boldsymbol{r}_{\perp} +r_{\mathrm{F}}^2 \boldsymbol{k}\right|^2 \nonumber\\
&+i \boldsymbol{k} \cdot \boldsymbol{r}_{\perp}-i \frac{r_{\mathrm{F}}^2}{2} k^2 .
\end{align}
By using $\int d \boldsymbol{r} e^{i a|\boldsymbol{r}|^2}=i \pi / a$ we can perform $\boldsymbol{r}$-integral to obtain
  \begin{align}
  \frac{\phiobs_1}{\phiobs_0} 
  &= \int \frac{d^2\bm{k}}{(2\pi)^2} i\frac{e^{-ir^2_\text{F}k^2/2-1}}{r^2_\text{F}k^2/2} \tilde{\kappa}(\bm{k})e^{i\bm{k \cdot r}_\perp} \nonumber\\
  &= \int \frac{d^2\bm{k}}{(2\pi)^2} \tilde{G}(\bm{k}) \tilde{\kappa}(\bm{k})e^{i\bm{k} \cdot \bm{r}_\perp}, \label{eq:11}
  \end{align}
  where we define 
  \begin{equation}
  \tilde{G}(\bm{k}) \equiv i\frac{e^{-ir^2_\text{F}k^2/2-1}}
  {r^2_\text{F}k^2/2}.
  \end{equation}
  This is a general expression of the relation between the distortion caused by wave optics gravitational lensing and the mass distribution of a lens object in Fourier space.  Now we consider the expression in real space. We use
  \begin{equation}
  G(\bm{r}) = \int \frac{d^2\bm{k}}{2\pi} \tilde{G}(\bm{k})e^{i\bm{k} \cdot \bm{r}}
  = \int \frac{kdk}{2\pi} \tilde{G}(k)J_0(kr), \label{eq:hankel}
  \end{equation}
  where 
  \begin{equation}
  J_0(x) = \frac{1}{2\pi} \int^{2\pi}_0 e^{ix\cos \phi},
  \end{equation}
  is the Bessel function. Equation (\ref{eq:hankel}) is valid when $G(\bm{r})$ is spherical symmetric, $|\bm{k}| = k$ and $\kappa(\bm{k}) =\kappa(k)$. It is called the Hankel transform. Equation (\ref{eq:11}) can be rewritten as 
  \begin{align}
  \frac{\phiobs_1}{\phiobs_0} 
  &= \int d^2\bm{r} G(\bm{r}) \int \frac{kdk}{2\pi} \tilde{\kappa}(k) J_0(k(r_{\perp}-r)) \nonumber\\
  &= \int d^2\bm{r} \kappa(\bm{r}_\perp-\bm{r}) G(\bm{r}) \nonumber\\
  &= \int d^2\bm{r} \kappa(\bm{r}) G(\bm{r}-\bm{r}_\perp), \label{eq:lastborn}
  \end{align}    
   where
   \begin{align}
   G(\bm{r}) 
   &= \int \frac{d^2\bm{k}}{(2\pi)^2} i \frac{e^{i r^2_\text{F} k^2 / 2}-1}{r^2_\text{F} k^2 / 2} e^{i\bm{k} \cdot \bm{r}} \nonumber\\
   &= -\frac{i}{2\pi r^2_\text{F}} \Gamma \left(0, -\frac{r^2}{2 r^2_\text{F}}\right) \nonumber\\
   &= -\frac{i}{2\pi r^2_\text{F}} \left\{ \text{Ei} \left(\frac{r^2}{2 r^2_\text{F}}\right) - i \pi \right\}.
   \label{eq:eq04}
   \end{align}
  Equation (\ref{eq:lastborn}) shows that the distortion of the waveform of the gravitational waves and the mass distribution of a lens object are connected with each other by the convolution in the Born approximation. Fig.~\ref{fig:epsart} shows the kernel function of the convolution, $G(\bm{r})$. It has a damped oscillation in the region $r > r_\text{F}$. This means that equation (\ref{eq:lastborn}) is equal to the smoothing with the Fresnel scale $r_\text{F}$. It can also be interpreted that the source effectively has a size of $r_\text{F}$ due to the wave optics effect.

   We note that \citet{2005A&A...438L...5T} derived an equation similar to equation (\ref{eq:lastborn}), without acknowledging that it is essentially the smoothed convergence. We also note that \citet{2021PhRvD.104f3001C} derived a similar equation but only for $\bm{r}_\perp=\bm{0}$.
   
   \subsection{The derivation of the post Born terms}
   Next, we derive the higher order terms of the Born approximation. While it is very difficult to derive the post Born terms directly, we argue that it can easily be derived by expanding the exponent containing the lens potential. Specifically, we start with the exact analytical form of the amplification factor for a single lens plane calculated by using the path integral \cite{1999PThPS.133..137N} 
   \begin{equation}
   \frac{\phiobs}{\phiobs_0}
   = \frac{1}{2\pi i r^2_\text{F}} \int d^2\bm{r} \exp \left[ i \frac{1}{2r^2_\text{F}} | \bm{r} - \bm{r}_\perp |^2 \right] \exp \left[ -\frac{i}{r^2_\text{F}} \psi (\bm{r}) \right] .
   \end{equation}
   We expand the exponential term  $\exp(-i\psi(\bm{r})/r^2_\text{F})$ around $\bm{r} = \bm{r}_\perp$ to obtain
   \begin{multline}
   \frac{\phiobs}{\phiobs_0}
   =1 - \frac{1}{2\pi r^4_\text{F}} \int d^2\bm{r} \exp \left[ i\frac{1}{2r^2_\text{F}}|\bm{r}-\bm{r}_\perp|^2 \right] \psi(\bm{r}) \\
   + \frac{1}{4\pi r^6_\text{F}} \int d^2\bm{r} \exp \left[ i\frac{1}{2r^2_\text{F}}|\bm{r}-\bm{r}_\perp|^2 \right] \psi^2(\bm{r}) + \cdots .\label{eq:postborn_}
   \end{multline}
   The second term on the right side of this equation is equal to equation (\ref{eq:fresnelborn}). The third term is the post Born term. The higher order terms can obviously be obtained also from this expansion.

   \section{The case of a point mass lens}
   
   We describe the explicit form of the exact analytical form of the amplification factor as well as derive expressions and behaviors of the Born and post-Born terms when the lens model is given by a point mass lens with the lens potential
   \begin{equation}
   \psi(\bm{r}) =r^2_\text{Ein} \log \left( \frac{|\bm{r}|}{r_\perp} \right),
   \end{equation}
   which satisfies the condition of $\psi(\bm{r}_\perp) = 0$  discussed in Sec.~\ref{sec:redef_phase}.
   While the Born approximation is not needed in most cases for a point mass lens because the exact analytical form of the amplification factor is known, we expect that the detailed analysis of the Born approximation for a point mass lens will lead to a better understanding of the Born approximation itself and will provide a useful guidance for future studies of the Born approximation for other lens mass models. In what follows, we use dimensionless parameters defined as
    \begin{equation}
   y = \frac{r_\perp}{r_\text{Ein}}, \qquad  w = \frac{r^2_\text{Ein}}{r^2_\text{F}},
   \end{equation}
   where the Einstein radius $r_\mathrm{Ein}$ is given by
   \begin{equation}
   r_\text{Ein} = \sqrt{\frac{4 G M}{c^2} \frac{\chi\left(\chi_{\mathrm{s}}-\chi\right)}{a \chi_{\mathrm{s}}}}.
   \end{equation}
   The scale factor $a$ corresponds to that at the lens redshift. $y$ is the impact parameter between the source and lens object on the source plane normalized by the Einstein radius. The Einstein radius represents roughly the effective size of the lens object. The parameter $w$ represents the degree to which the geometrical optics effect or the wave optics effect is dominant. When $w$ is smaller than 1, the wave optics effect dominates, and when $w$ is larger than 1, the geometrical optics effect dominates. Or, if the mass of the lens object is fixed, it is seen that $w \propto f$. Therefore $w$ can also be seen as a normalized frequency. 
   
   \subsection{Exact analytical form}
   The amplification factor for a point mass lens is already calculated as \cite{1986ApJ...307...30D}
   \begin{align}
   1+\frac{\delta \phiobs}{\phiobs_0} =
   &\exp \left(\frac{\pi w}{4}+i \frac{w}{2}\left(\ln \left(\frac{w}{2}\right)-\left(x_m-y\right)^2 +2 \ln x_m \right)\right) \nonumber\\
   & \times \Gamma\left(1-i \frac{w}{2}\right){}_1 F_1\left(i \frac{w}{2}, 1 ; \frac{w y^2}{2}\right), 
    \label{eq:analytic_point}
   \end{align}
   where
   \begin{equation}
   x_m=\frac{y+\sqrt{y^2+4}}{2},
   \end{equation}
   $\Gamma$ is the gamma function, and ${}_1 F_1$ is the confluent hypergeometric function. This exact analytical form is derived under the thin lens approximation, which is valid when the size of a lens object is much smaller than the distance to the lens object \cite{2005PhRvD..72d3001S}. In general, the distance to the lens object is cosmological and hence we can assume that the thin lens approximation and the exact analytical form (\ref{eq:analytic_point}) is always valid. In the geometrical optics limit $w \rightarrow \infty$ \cite{2003ApJ...595.1039T}
\begin{align}
1 + \frac{\delta\phiobs}{\phiobs} & =\left|\mu_{+}\right|^{1 / 2} \exp \left[i w\left(\frac{1}{2}\left(p_{+}-y\right)^2-\log \left|p_{+}\right|\right)\right] \nonumber\\
& -i\left|\mu_{-}\right|^{1 / 2} \exp \left[i w\left(\frac{1}{2}\left(p_{-}-y\right)^2-\log \left|p_{-}\right|\right)\right],
\end{align}
   where the magnification of each image $\mu_{ \pm}=1 / 2 \pm(y^2+2) /(2 y \sqrt{y^2+4})$ and $p_{ \pm}=(1 / 2)(y \pm \sqrt{y^2+4})$.
   \subsection{Born term}
   
     When the lens model is a point mass lens,  we can obtain the Born term from equation (\ref{eq:lastborn}) as
   \begin{equation}
   \frac{\phiobs_1}{\phiobs_0}
   = \frac{i}{2}w \left\{ \text{Ci} \left( \frac{wy^2}{2} \right) + i \text{Si} \left( \frac{wy^2}{2} \right) -i \frac{\pi}{2} \right\} .
   \end{equation}
   For large $y$ and large $w$, the behavior of $\phiobs_1/\phiobs_0$ is derived as
   \begin{equation}
   \frac{\phiobs_1}{\phiobs_0} \propto \frac{1}{y^2}+\frac{1}{wy^4}.
   \end{equation}
   We ignore the phase for simplicity. In geometric optic limit $w \rightarrow \infty$, $y$-dependence is derived as
   \begin{equation}
   \frac{\phiobs_1}{\phiobs_0} \propto \frac{1}{y^2}.
   \end{equation}
   This means that the distortion due to the gravitational lensing effect  reaches to the constant value $1/y^2$ in the geometrical optics limit. The point where the value starts to approach the asymptotic value, i.e., where the $w$-dependence disappears, is $wy^4 \gg y^2$, or equivalently $w \gg y^{-2}$.
   
   \begin{figure}[tbp]
   \includegraphics[width=9cm]{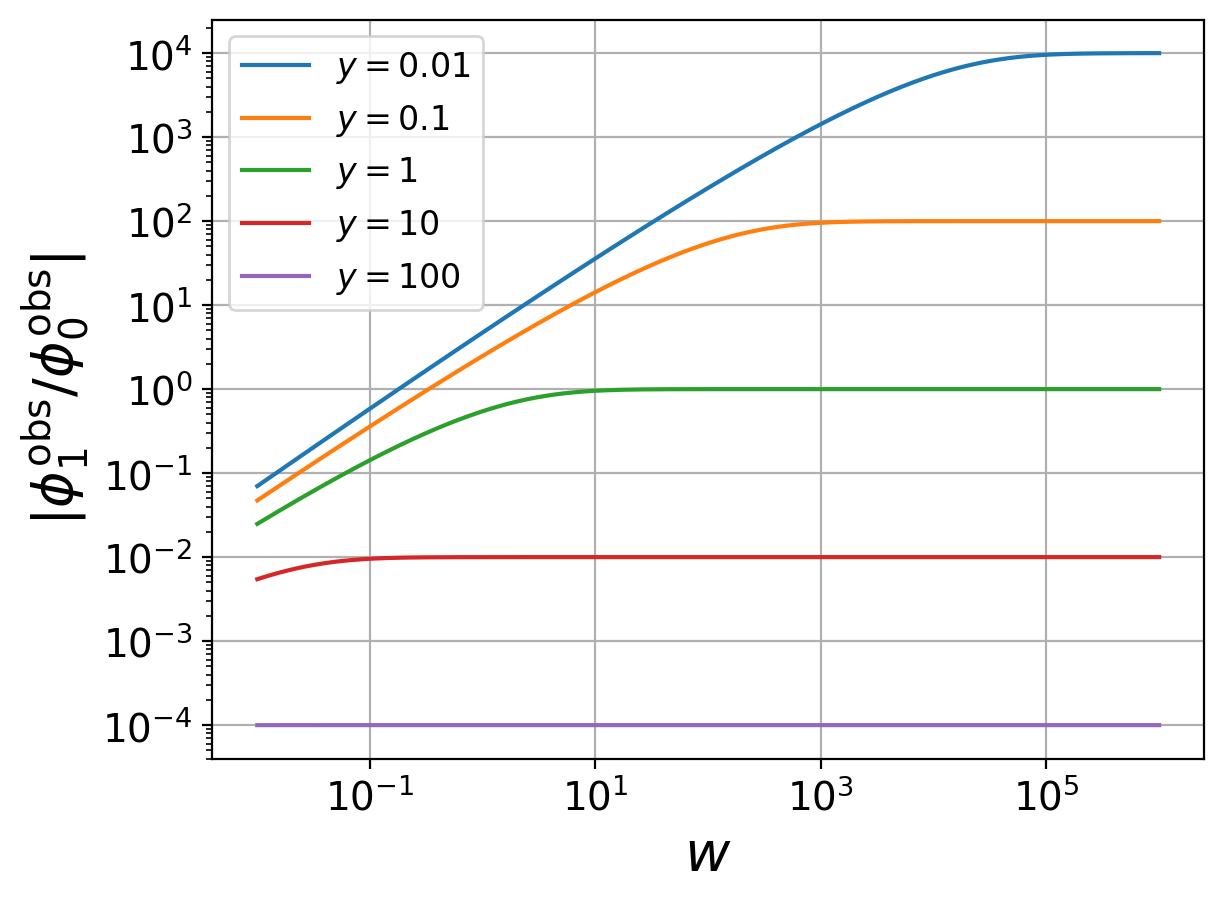}
   \caption{\label{fig:F_abs_w_Born} The absolute value of $\phi_1/\phi_0$ as a function of the normalized frequency $w$. The normalized impact parameter $y$ is changed from $10^{-2}$ to $10^{2}$ from top to bottom lines.}
   \end{figure} 
  
   Fig.~\ref{fig:F_abs_w_Born} shows the $w$-dependence of the absolute value of the distortion in the Born approximation. At large $w$, as $y$ increases by ten fold, the absolute value of the Born term also increases by a hundred fold. This represents  $\phiobs_1 / \phiobs_0 \propto 1 / y^2$ in the geometrical optics limit. Furthermore, the beginning of asymptotic follows $w \sim y^{-2}$. We can see that the difference between the exact analytical form and the Born approximation is also proportional to $1/y^2$. Therefore, as parameter $y$ increases, the difference decreases for large $w$.

 \begin{figure*}[tbp]
    \centering
    \begin{minipage}{0.45\textwidth}
        \centering
        \includegraphics[width=\textwidth]{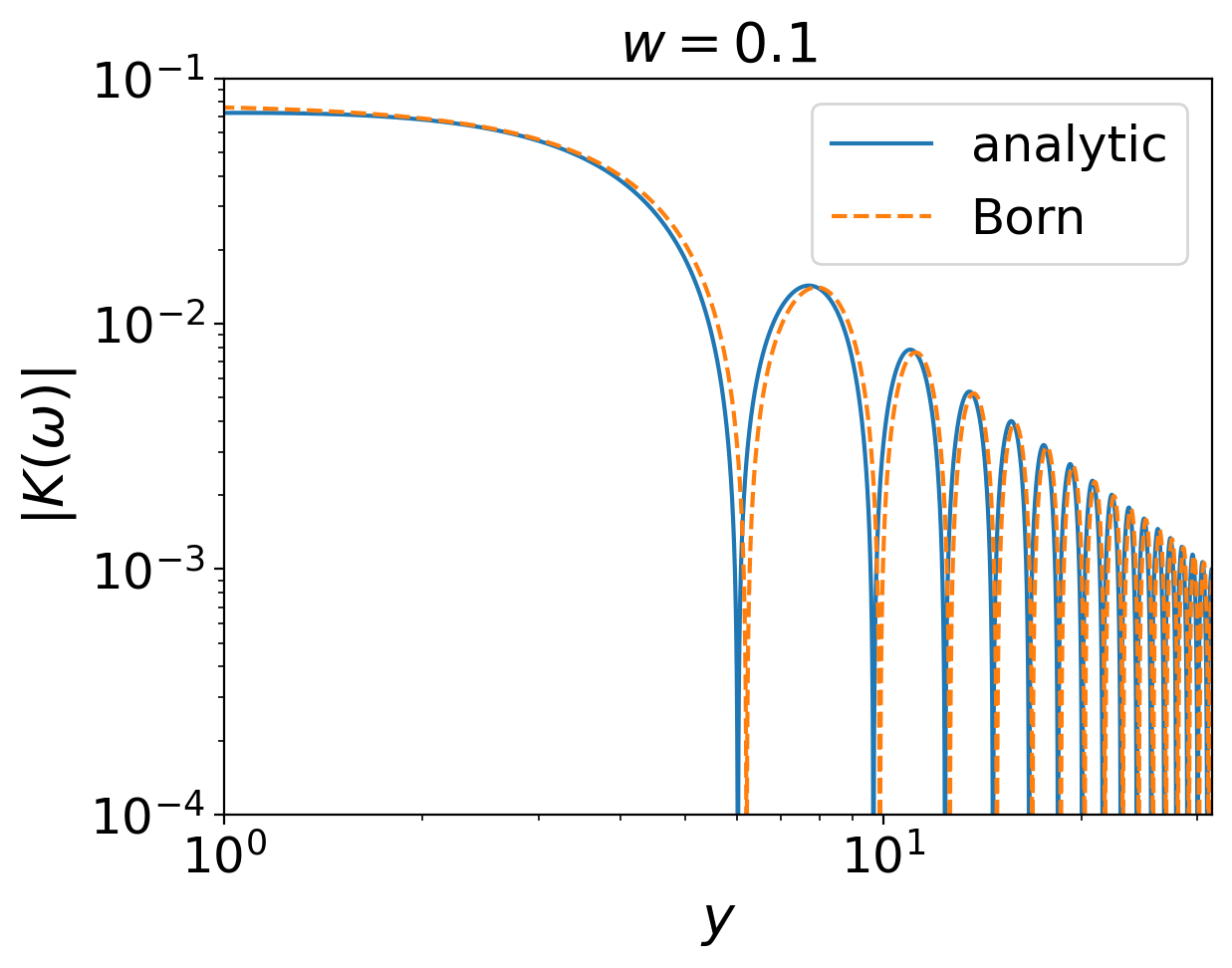}
    \end{minipage}
    \hfill
    \begin{minipage}{0.45\textwidth}
        \centering
        \includegraphics[width=\textwidth]{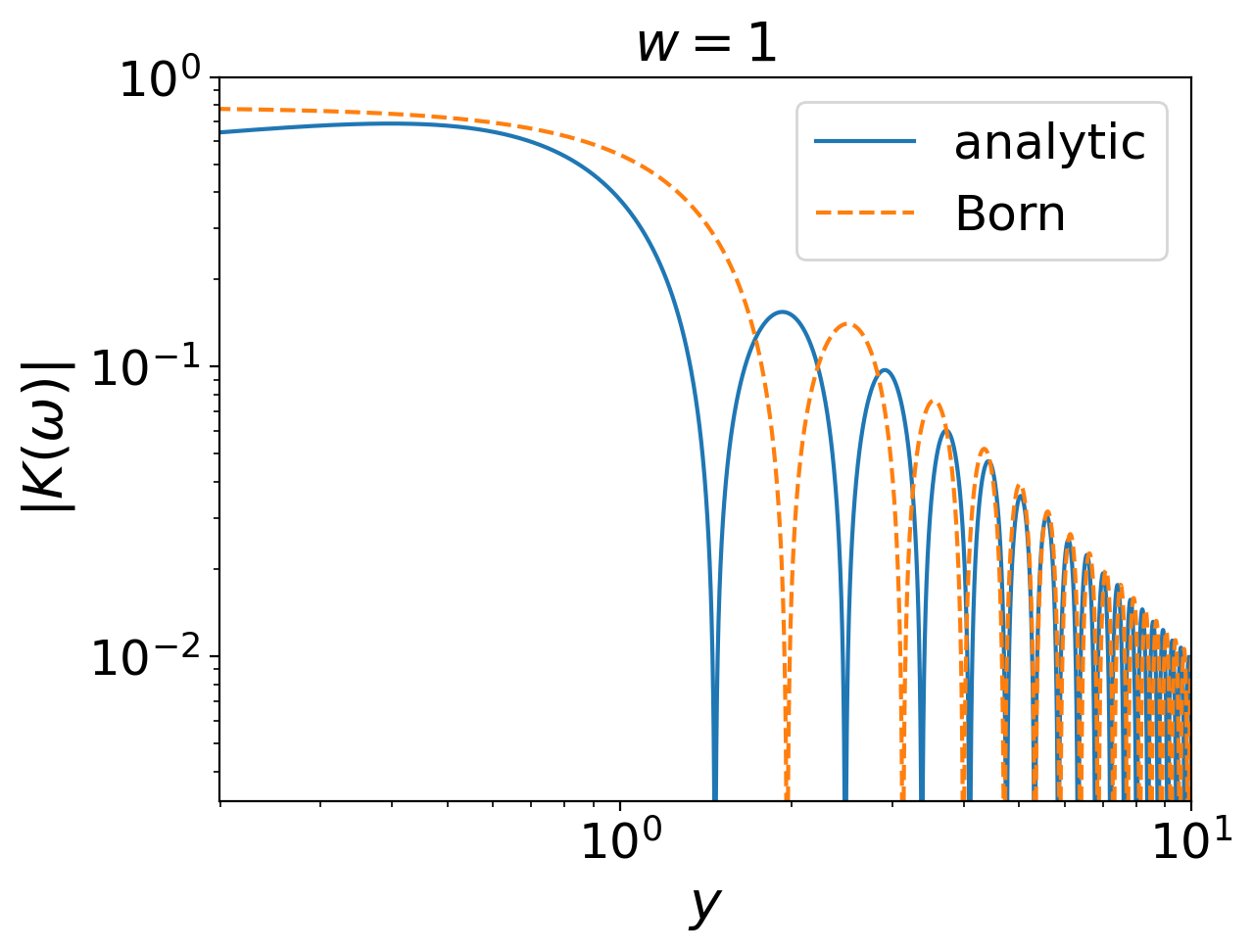}
    \end{minipage}
    \caption{The comparison of the absolute values of $K(\omega)$ defined in equation (48) between the exact analytical form ({\it solid}) and the Born approximation ({\it dashed}). The left and right panels show results for $w=0.1$ and $1$, respectively.}
    \label{fig:fig3}
\end{figure*}

   \subsection{Post-Born terms}
   
   For the case of a point mass lens, we can obtain the post-Born term from equation (\ref{eq:postborn_}) as
   \begin{equation}
   \frac{\phiobs_2}{\phiobs_0}
   = \frac{i}{4\pi} w^3 y^2 e^{\frac{1}{2}iwy^2} \int dz z \log{z}^2 J_0(wy^2 z) e^{\frac{1}{2}iwy^2 z^2},
   \label{eq:second}
   \end{equation}
   With a crude approximation ignoring the oscillating part of the Bessel function as well as the exponential function, we obtain
 \begin{align}
 \frac{\tilde{\phi}^{\text {obs }}_2}{\tilde{\phi}^{\text {obs }}} 
 &= \frac{i}{4 \pi} w^3 y^2 e^{\frac{1}{2} i w y^2} \int_0^{\infty} dz\,z \log ^2 z J_0\left(w y^2 z\right) e^{\frac{1}{2} i w y^2 z^2}\nonumber\\
 &\simeq \frac{i}{4 \pi} w^3 y^2  \int_0^{\frac{1}{w y^2}} dz\,z\log ^2 z\nonumber\\
 &\propto \frac{w}{y^2}\left[1+2\log\left(\frac{1}{w y^2}\right)\left\{\log\left(\frac{1}{w y^2}\right)-1\right\}\right]\nonumber\\
 &\sim \frac{w}{y^2}.
 \label{eq:eq05}
 \end{align}
 
 We apply this method to the $n$-th Born term to obtain
 \begin{align}
\frac{\tilde{\phi}^{\text {obs }}_n}{\tilde{\phi}_{\text {obs }}} 
 =& \frac{(-1)^n i^{n-1}}{2 \pi n!} w^{n+1} y^2 e^{\frac{1}{2} i w y^2} \nonumber\\
&\times \int_0^{\infty} dz\,z \log ^n z J_0\left(w y^2 z\right) e^{\frac{1}{2} i w y^2 z^2} \nonumber\\
&\propto \frac{w^{n-1}}{y^2},
  \end{align}
 when $n$ is integer larger than one.
   
   \section{The accuracy of the Born approximation}
   
       \begin{figure*}[tbp]
    \centering
    \begin{minipage}{0.45\textwidth}
        \centering
        \includegraphics[width=\textwidth]{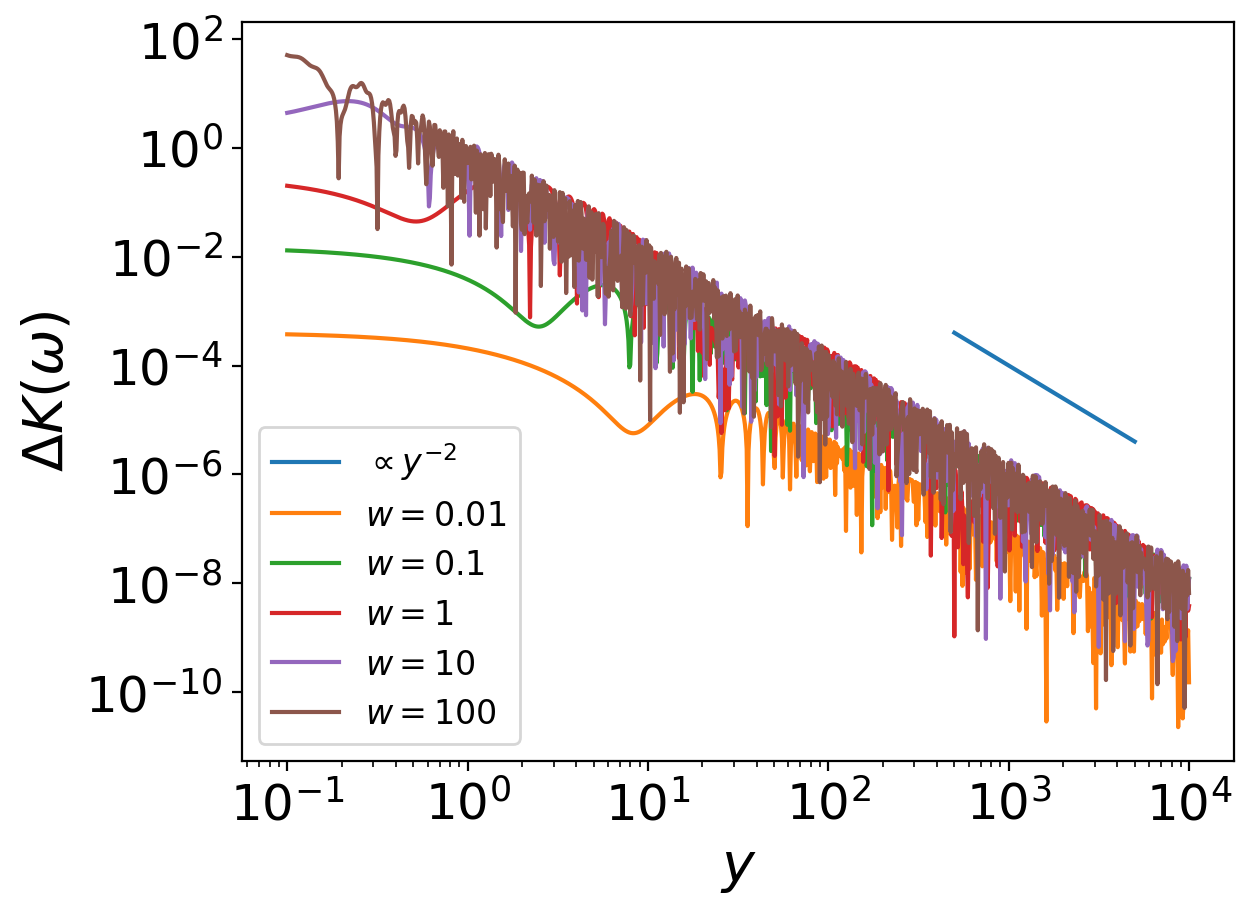}
    \end{minipage}
    \hfill
    \begin{minipage}{0.45\textwidth}
        \centering
        \includegraphics[width=\textwidth]{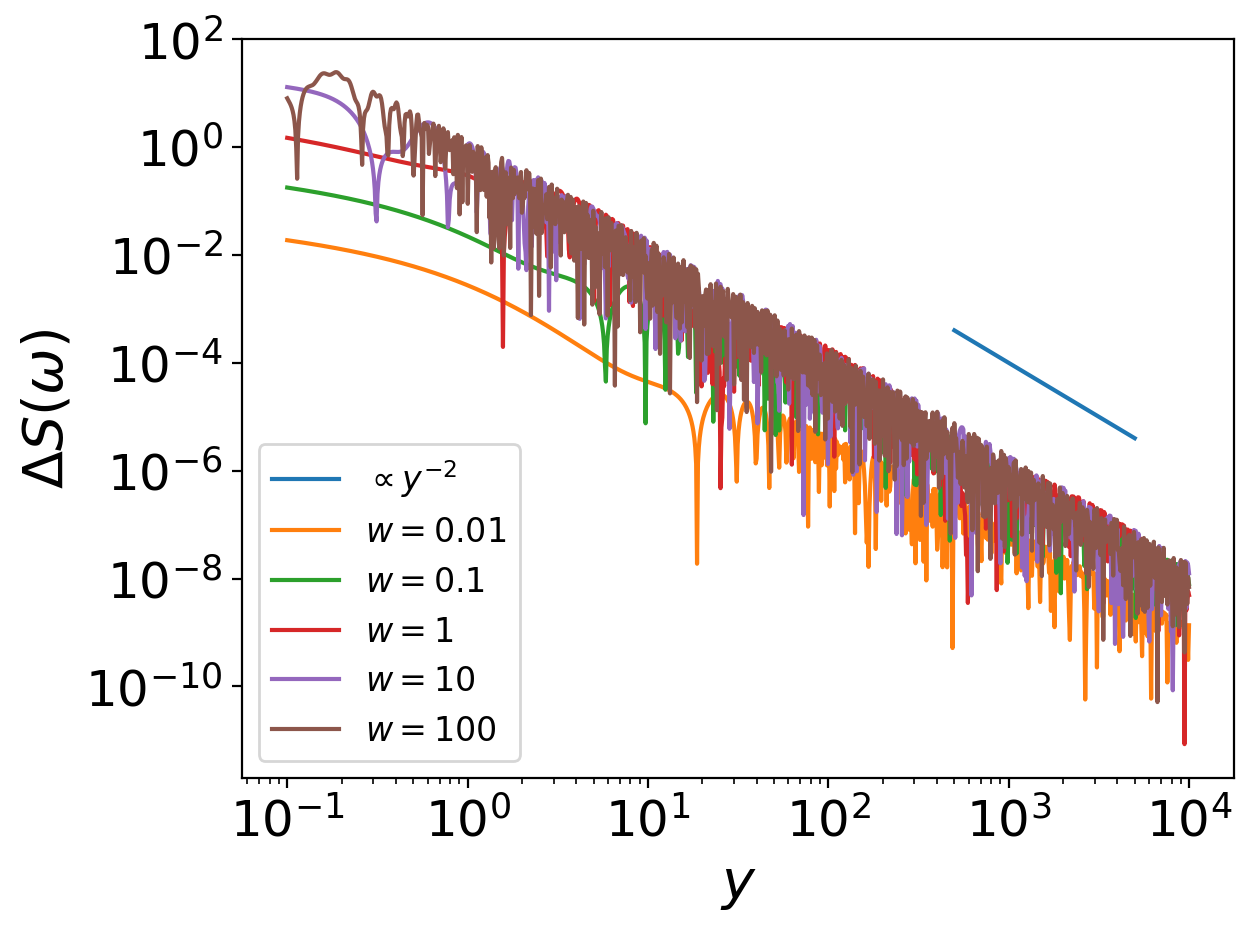}
    \end{minipage}
    \caption{The differences of the real ({\it left}) and imaginary ({\it right}) parts between the exact analytical form and the Born approximations. We show results for different values of $w$ between $10^{-2}$ and $10^2$.}
    \label{fig:fig4}
\end{figure*}

   \subsection{Amplitude and phase}
   
   Next, we investigated the accuracy of the Born approximation. First, we summarize the relationship between real parts and imaginary parts. Let us define $K$ and $S$ as (\cite{104210}, Ch.~17)
   \begin{gather}
   K(\omega) = \text{Re}\left[ \frac{\phiobs_1}{\phiobs_0} \right], \\
   S(\omega) = \text{Im}\left[ \frac{\phiobs_1}{\phiobs_0} \right].
   \end{gather}
   
   In the Born approximation, $\phiobs_1/\phiobs_0$ can be rewritten as 
   \begin{equation}
   \frac{\phiobs_1}{\phiobs_0} \simeq (1+K(\omega))\exp[iS(\omega)].
   \end{equation}

   Considering a point mass lens, we compare the exact analytical form with approximate solution. Specifically, we calculate
   \begin{gather}
   \Delta K \equiv  K_\text{exi}(\omega)-  K_\text{Born}(\omega), \\
   \Delta S \equiv  S_\text{exi}(\omega) -  S_\text{Born}(\omega).
   \end{gather}
    
    \subsection{Result}
    
    First, in Fig.~\ref{fig:fig3} we show examples of the behavioral differences between the exact analytical form and the Born approximation. When $w$ is small, which correspond to the wave optic regime, the exact analytical form and Born approximation are very similar. When $w$ increase, however, the behavior of two solutions begin to deviate at small $y$, although they behave similarly at large $y$.
   
    Fig.~\ref{fig:fig4} shows $y$-dependence of the differences between the exact analytical form and the Born approximation. It is shown that as $w$ increases, the difference also increases. However, the increase stops when $w>1$. This is because exact analytical expression and the Born approximation have the same dependence of the parameter $y$ in the geometrical optics limit. We also find that the differences decrease as $y^{-2}$ as $y$ increases.
 
\begin{figure}[tbp]
   \includegraphics[width=9cm]{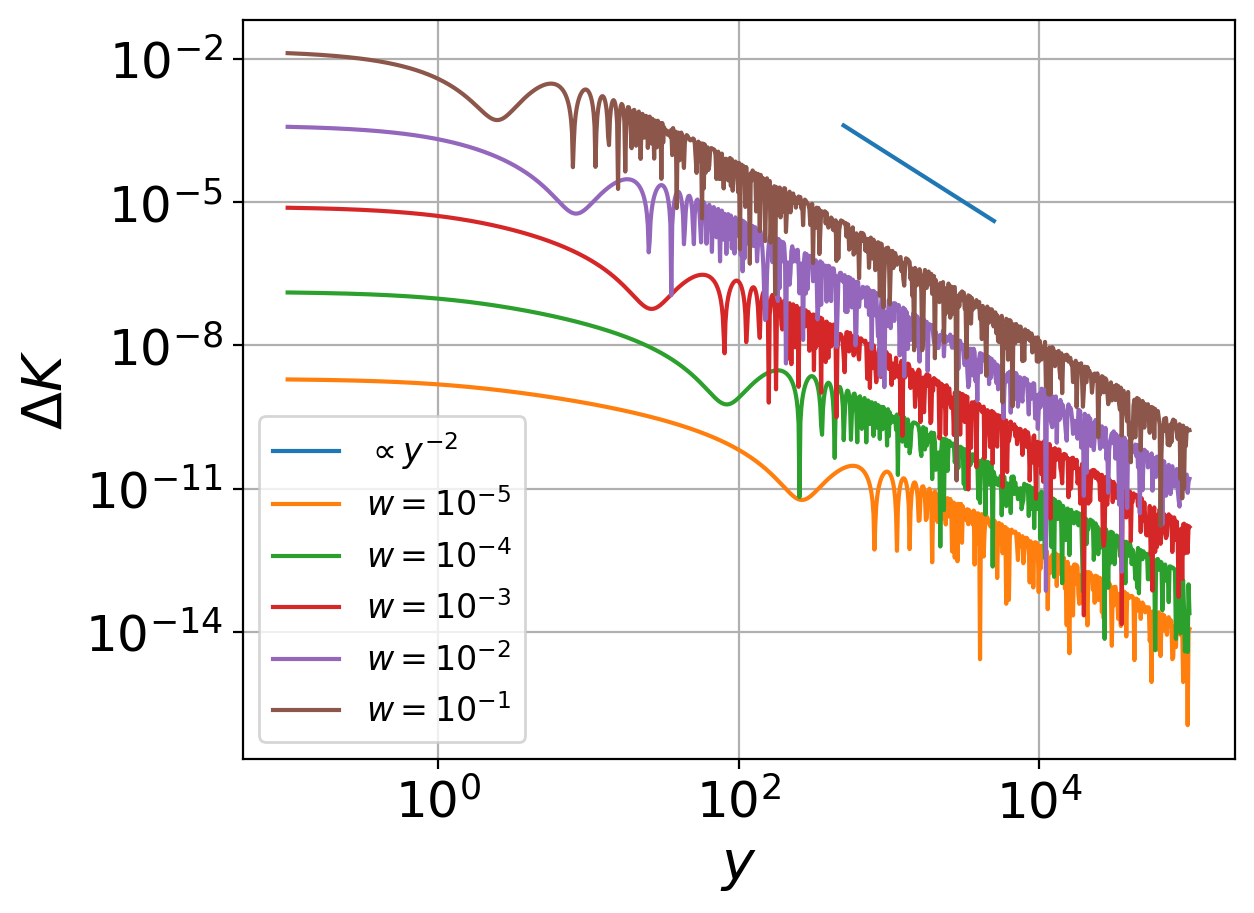}
   \caption{\label{fig:deltak_y_smallw} The difference of the real part between the exact analytical expression and the Born approximation where the parameter $w$ is small, $10^{-5} < w < 10^{-1}$. }
   \end{figure} 
     
Fig.~\ref{fig:deltak_y_smallw} shows the differences  when the parameter $w$ is small. From this Figure, for large $y$, we find that as $w$ increases by one digit, the difference value also increases by one digit, indicating that the difference is proportional to $w$. To summarize, we find
 \begin{equation}
\Delta K(\omega) \propto \frac{w}{y^2}, \qquad \Delta S(\omega) \propto \frac{w}{y^2}.
\label{eq:eq06}
 \end{equation}
   The Born approximation is an approximation that takes up to the first-order term of the successive expansion, so the difference between the Born approximation and the exact analytical form is dominated by the second term of the successive expansion. In equation (\ref{eq:eq05}), we derive that the second term should behave as $\propto w/y^2$, which is consistent with our numerical result summarized in equation (\ref{eq:eq06}).

\section{Conclusion}  
We have investigated the property of the Born approximation and its accuracy in wave optics gravitational lensing in detail for the case of a single lens plane. We have shown that,  under the Born approximation, the distortion of gravitational waves caused by wave optics gravitational lensing is in general connected with the mass distribution of a lens object by a convolution integral with a kernel whose size is characterised by the Fresnel scale. The behavior of the kernel indicates that the wave optics effect can be interpreted as a smoothing over the Fresnel scale. This result is consistent with \citet{2020ApJ...901...58O}, in which the similar smoothing effect is discussed in the context of weak lensing by the large-scale structure rather than a single lens plane. We have also shown that post Born higher order terms can be evaluated by the Taylor expansion of an exponential function containing the lens potential.

We have checked the validity and accuracy of the Born approximation focusing on a point mass lens for which the exact solution of the amplification factor is known. By using the dimensionless parameter $y$ that represents the normalized parameter between the lens and the source object and the dimensionless parameter $w$ that represents the normalized frequency, we have found that the $n$-th term of the Born approximation is proportional to $y^{-2}w^{n-1}$. This result implies that the Born approximation is valid when the parameter $w$ is smaller than 1. In this case, the accuracy of the Born approximation is proportional to $wy^{-2}$. We have verified this behavior both analytically and numerically.

The methodology developed in this paper is useful for studying the Born approximations of other lens models. We will explore this application in future work.

\begin{acknowledgments}
We thank the anonymous referee for careful reading of the manuscript and for useful comments.
This work was supported by JSPS KAKENHI Grant Numbers JP22K18720, JP22K21349, and JP20H05856.
\end{acknowledgments}

\bibliography{ref}

\end{document}